\documentclass[aps,preprint,amssymb,12pt,floatfix]{revtex4}
\usepackage{subfigure}
\usepackage{floatflt}
\usepackage{color}
\usepackage{pifont}
\input{epsf.tex}
\usepackage{rotating}

\begin{document}

\title{Quasi-one dimensional fluids that exhibit higher dimensional
behavior}

\author{SILVINA M. GATICA\footnote{Corresponding author}}

\affiliation{Department of Physics and Astronomy, Howard
University, Washington, DC, 20059, USA\\
sgatica@howard.edu}

\author{M. MERCEDES CALBI}

\affiliation{Department of Physics, Southern Illinois University,
Carbondale, IL
USA}

\author{GEORGE STAN}

\affiliation{Department of Chemistry, University of Cincinnati, Cincinnati,
OH
45221, USA}

\author{R. ANDREEA TRASCA}

\affiliation{Dieffenbachstr. 58A, 10967 Berlin,
Germany}

\author{MILTON W. COLE$^*$}

\affiliation{Department of Physics, Penn State University, University Park,
PA
16802, USA\\ 
miltoncole@aol.com}

\baselineskip 14pt

\begin{abstract}
Fluids confined within narrow channels exhibit a variety of phases and
phase 
transitions associated with their reduced dimensionality. In this review
paper,
we illustrate the crossover from quasi-one dimensional to higher effective
dimensionality behavior of fluids adsorbed within different carbon
nanotubes
geometries. In the single nanotube geometry, no phase transitions can occur
at
finite temperature. Instead, we identify a crossover from a quasi-one
dimensional to a two dimensional behavior of the adsorbate. In bundles of
nanotubes, phase transitions at finite temperature arise from the
transverse
coupling of interactions between channels.
\end{abstract}

\maketitle

\section{Introduction}

One of the most interesting topics within modern condensed matter physics is
that of phenomena in reduced dimensionality, resulting from some degree of
spatial localization of the particles comprising a system.\cite{dimen} For example,
chemists, materials scientists and physicists have created and explored numerous
physical systems in which atoms and molecules are confined within quasi-one
dimensional (Q1D) environments. The variety of these systems is remarkable, such
as the peapod geometry, \emph{i.e.}, a line of buckyballs within a carbon
nanotube,\cite{bucky1,bucky2} fluids within artificial materials created by 
templating \cite{templ1,templ2}
 and Q1D optical lattices created by laser fields.\cite{optlatt}
Unfortunately, as far as
we know, there exists no comprehensive review of this general problem, although
many relevant subfields have been summarized.\cite{rev1,rev2,rev3,rev4} The present paper
addresses a small subset of this exciting research field. Specifically, we
consider problems involving fluids, both classical and quantum, confined within
Q1D channels, the focus of our group's research during  the last decade.\cite{revlowT} 

Here, the term Q1D refers to a system in which particles move in an external
potential field $V(\textbf{r})$ which is either constant or slowly varying in
\emph{one} direction ($z$), while $V$(\textbf{r}) is strongly localizing in the
two other (transverse) directions. In the case of quantum particles, for which
the transverse spectrum is discrete, one expects that the corresponding degrees
of freedom are frozen out at low temperature ($T$); transverse excitation does
occur at higher $T$, as determined by the gaps in the transverse spectrum of
states. This plausible expectation is borne out in some cases, but we shall see
that there can be dramatic consequences of the transverse degrees of freedom in
other cases, even at low $T$.

One of the many exciting aspects of strictly 1D physics is its susceptibility to
weak perturbations. The reason for this behavior arises from the fact that (in
all practical situations)\cite{practsit} no phase transition can exist in a purely 1D
system at any finite $T$, even though the ground state may exhibit
symmetry-breaking order. Thus, there do exist transitions at $T$ identically
equal to zero. A familiar example is the 1D Ising model, for which the
correlation length and susceptibility both diverge as $T$ approaches 0.

Since the purely 1D system has no finite $T$ transition, what brings about more
interesting behavior in the Q1D case? As discussed below, the difference can
arise from considering a set of parallel 1D systems which are weakly coupled.
Alternatively, the behavior can happen because the system is only 1D insofar as
the transverse dimensions are finite, unlike the length in the z direction, L,
which achieves the thermodynamic limit.....but the transverse dimensions
\emph{are} large enough to have an observable effect (e.g. low energy gaps). 
A third, more surprising,
origin of interesting phenomena is when the system is a collection of
noninteracting 1D systems, with "quenched" heterogeneity, which are coupled to a particle
bath, so they possess a common chemical potential.

The next section discusses the conceptually and computationally simplest case of
a Q1D system: a low density, noninteracting gas within a single channel; then,
the solution of the one-particle Schr\"odinger equation determines the physical
behavior. Section 3 considers the case of many \emph{interacting} particles
within a single channel; such a problem is often used as a model for fluids
within regular or irregular porous materials. Section 4 considers the problem of
Q1D channels containing fluids that interact with one another as well as with
fluids occupying other channels. 

\begin{figure}[htbt]
\subfigure[]{
\includegraphics[width=2.5in,height=2.25in]{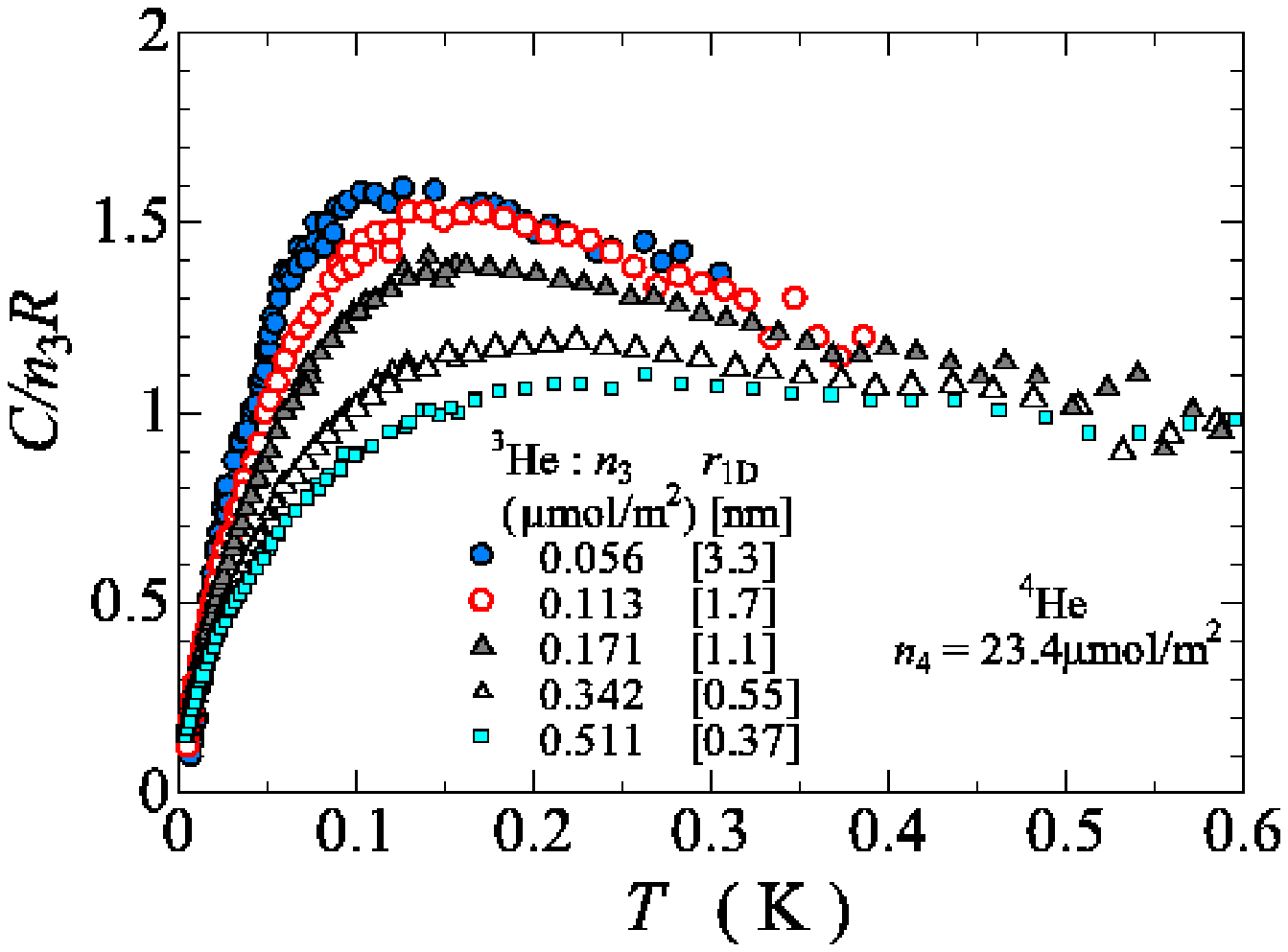}}
\subfigure[]{
\includegraphics[width=2.5in,height=1.25in]{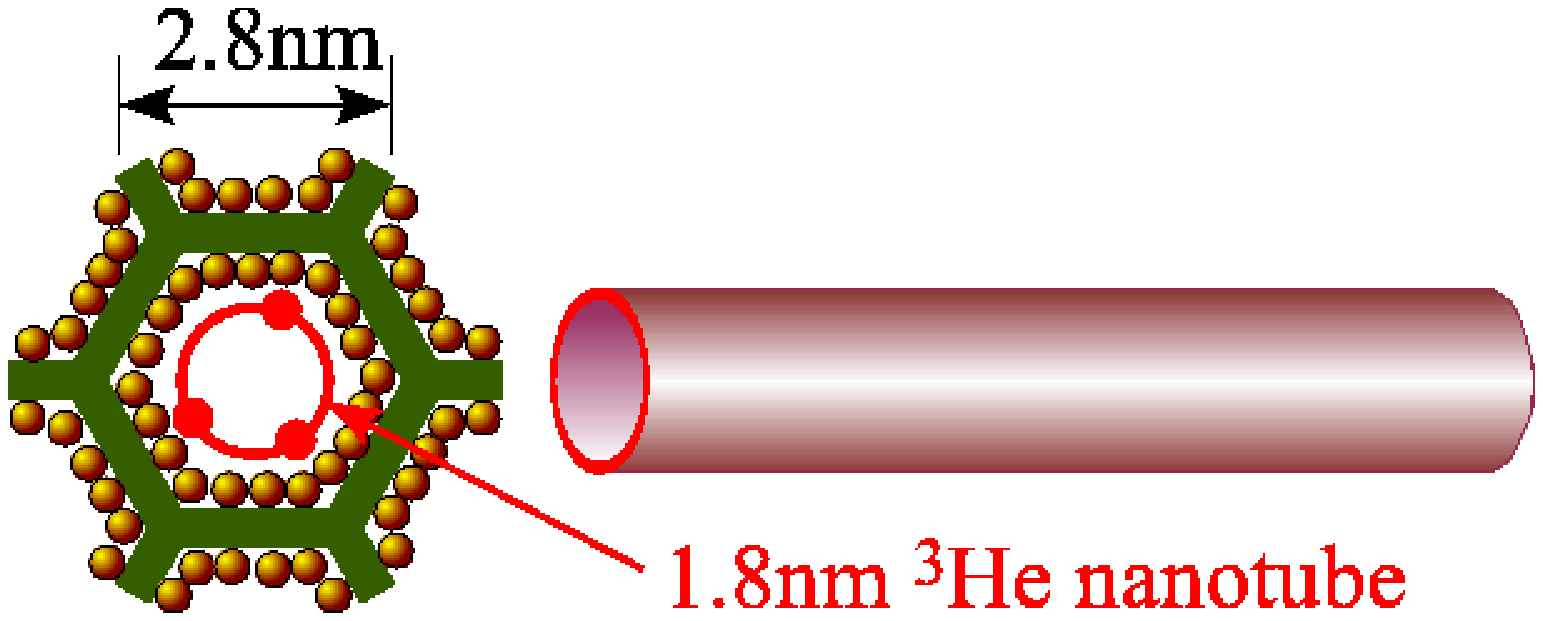}}
\caption{Left: Experimental heat capacity for $^3$He at various
densities, inside $^4$He-coated FSM-16, from Ref.~\cite{templ2}.
Right: schematic depiction of the geometry,
from Ref.~\cite{matsushita}. The tube represents the environment
experienced by a $^3$He atom.}
\label{fig:exp_C}
\end{figure}

\section{Low density gas in a single Q1D channel}

Our first problem is conceptually and calculationally simple: a low density gas
confined within a Q1D geometry; nevertheless, it provides interesting, and
sometimes surprising, results. One example that has been studied extensively is
that of a gas inside single carbon nanotubes, \emph{e.g.} quantum fluids or the
``peapod'' case of C$_{60}$ molecules.\cite{bucky2} Another is the case of
quantum gases
inside templated regular pores.\cite{templ2} A third example is the so-called
groove
region between two nanotubes, \emph{e.g.} on the outside of a bundle of
nanotubes.\cite{gr1,gr2,gr3} 

A \emph{classical} noninteracting gas has a kinetic energy per particle of (3/2)
$k_B T$ and a mean potential energy $\langle U \rangle$ determined by its
interaction with the environment. For the case of a particle localized near the
z axis within a channel, $\langle U \rangle = k_B T$, due to two transverse
directions of excitation. Hence, the classical  specific heat per particle is
$[C(T)/N]_{classical} =(5/2) k_B$. For a quantum gas, instead, the transverse
degrees of freedom are frozen out at low $T$, so one expects $[C(T)/N]_{quantum}
= (1/2) k_B$. The generalization to $D$ ``effective'' dimensions yields this
expression for the dimensionless specific heat, $C^*$, of a noninteracting
Boltzmann gas:
\begin{equation}
C* \equiv C/(Nk_B) = D/2 
\end{equation}

\begin{figure}[htbp]
\includegraphics[width=4.in,height=3.5in]{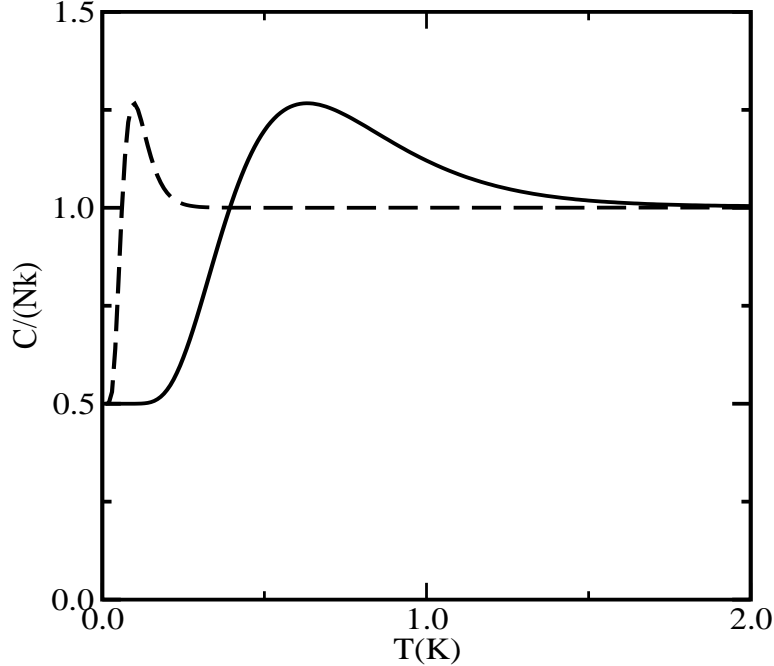}
\caption{Heat capacity for a low density gas of $^4$He atoms within a
single wall carbon nanotube, for cases $R$=0.8 nm (dashes) and 0.5 nm (full
curve). Quantum statistical effects are omitted from the calculation.
Adapted from Ref.~\cite{stan98surfsci}.}
\label{fig:theor_C}
\end{figure}

Fig. \ref{fig:exp_C} shows experimental results for $C^*$ in the case of
$^3$He inside of FSM-16, a material consisting of straight hexagonal pores of
cross-sectional distance of order 2 to 3 nm, pre-coated with a thin film of
$^4$He. At the low densities shown here, the $^3$He gas can be assumed to be
noninteracting, although interaction effects appear at higher
density.\cite{inteff} The
behavior observed inside FSM-16 can be understood by analogy to calculations in
Fig. \ref{fig:theor_C}, for a noninteracting gas of $^4$He within a carbon
nanotube. Note the overall similarity of these two figures.

In Fig. \ref{fig:theor_C},\cite{stan98surfsci} one observes that the low
$T$ limit is $C^* =1/2$,
corresponding to a 1D classical gas, as expected. At high $T$, instead, the
limit is $C^*=1$; this limit is interpreted as that of a 2D gas moving on
the
inner surface of the nanotube. The bump at intermediate $T$ is a general
property found for spectra, such as that of the rigid rotor,\cite{pathria}
for which the
inter-level spacing increases with quantum number. In the present case, the
relevant spectrum is that arising from the azimuthal kinetic energy, 
\begin{equation}
E_\theta = (\hbar \nu)^2/(2m \langle \rho \rangle^2) \equiv \nu^2 k_B
\Theta
\end{equation}
Here $\nu=0,\pm 1, \pm 2, \dots $ is the azimuthal quantum number and
$\rho$ is
the radial coordinate, while $\Theta$ is defined as a temperature
characteristic
of azimuthal excitation. The peak in the specific heat occurs at a
temperature
near 3 $\Theta$ for both radii considered in Fig. \ref{fig:theor_C}, so its
position serves as a benchmark from which one can determine the value of
$\langle \rho \rangle$. Note that inside a nanotube, the relevant value of
$\langle \rho \rangle$ is typically $R-\sigma$, where $\sigma$ is the
gas-surface hard core interaction length. The key qualitative difference
between
the behaviors seen in Figs. \ref{fig:exp_C} and \ref{fig:theor_C} is that
the
experimental data in Fig. \ref{fig:exp_C} plunge to $C^*=0$ as
$T\rightarrow0$. This is an effect of quantum degeneracy, manifested in
Nernst's
law, as found in explicit calculations which revise Fig. \ref{fig:theor_C}
by
taking quantum statistics into account.\cite{qstat}

\begin{figure}[htbp]
\includegraphics[width=4in]{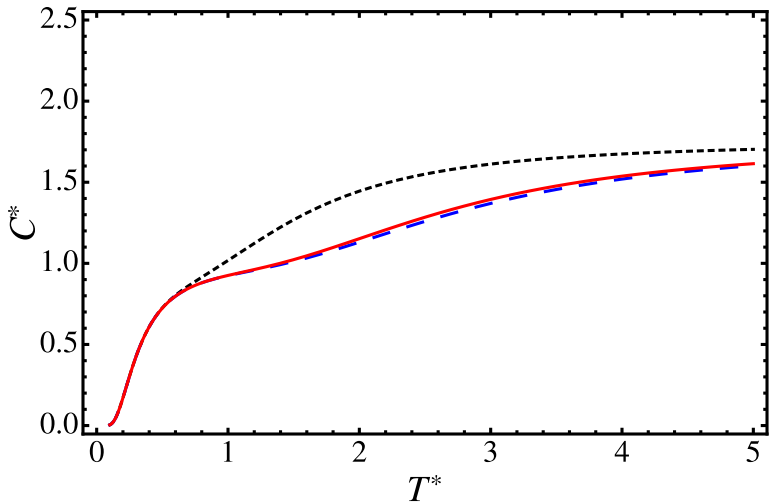}
\vspace*{8pt}
\caption{Reduced specific heat C* for the gases He (dashed-blue), H$_2$
(solid-red) and Ne (short-dashed-black) between two
nearly parallel nanotubes, each with $R = 0.7$ nm, and a divergence
half-angle $\gamma = 0.5$ degrees. From Ref.~\cite{dimen}.}
\label{fig:groove_C}
\end{figure}

There have been many theoretical studies of gas adsorption in the presence
of
nanotubes.\cite{rev1,rev2,rev3,rev4,revlowT} In most treatments of these
systems, one 
assumes that neighboring
tubes are parallel. In that case, there exists a region of space- the
so-called
``groove''- which is a 1D channel with a strongly attractive potential,
created
by the adjacent tubes. The adsorbed gas then exhibits Q1D behavior at low
$N$.
However, one can also inquire about the case when the tubes are not quite
parallel, but instead diverge, leaving a particularly attractive region
between
them, with a minimum potential energy ($V_0$) located at equilibrium
position
$r_0$. In a forthcoming study,\cite{dimen} we will report remarkable
results for this
geometry, as exemplified in Fig. \ref{fig:groove_C}. The low $T$ behavior
is
that of a gas in a 1D harmonic potential, $V \simeq V_0 + k_z (z-
z_0)^2/2$,
where $k_z$ is the force constant for particle motion parallel to the $z$
axis,
midway between and nearly parallel to the tubes' axes. We introduce the
characteristic temperature for this motion, $T_z \equiv \hbar 
(k_z/m)^{1/2}/k_B$ and a reduced temperature $T^* \equiv T/ T_z$. For
$T^* <<
1$, the specific heat is of Arrhenius form, as seen in the figure, while
for
$T^*\sim1$, $C^*\sim1$, the specific heat of a 2D gas. That behavior might
not
have been anticipated, at first glance, because for small divergence
half-angle
$\gamma$, one might have expected 1D behavior, i.e., $C^*=1/2$. Another
surprise
is the high $T^*$ limiting behavior, $T^* \rightarrow 7/4$. This peculiar
result
arises because the relevant particles' motions are those in the plane
perpendicular to the $x-z$ plane of the nanotubes, for which the potential
variation is unusual- proportional to $y^4$.

\section{Many interacting particles in a channel of finite width }

From the perspective of phase transitions, a channel of finite width is a
Q1D
system, so that no true thermodynamic singularities can occur. This means
that
many attempts to explore behavior in porous media with single channel
models
(such as cylindrical and slit pores) cannot accurately describe phase
transition
behavior that is seen in genuinely 3D porous media. Nevertheless, these
models
may provide good semiqualitative predictive power, sufficient for most
purposes
since the fully 3D geometry is not known. We have explored a variety of
such
models, with several different goals. These include assessing the accuracy
of
simplifying models, such as mean-field-theory (MFT) and the use of periodic
boundary conditions. Both of these approximations are suspect, at first
glance,
due to the important role of fluctuations in 1D statistical physics.

\begin{figure}[bt]
\includegraphics[width=5in,height=4in]{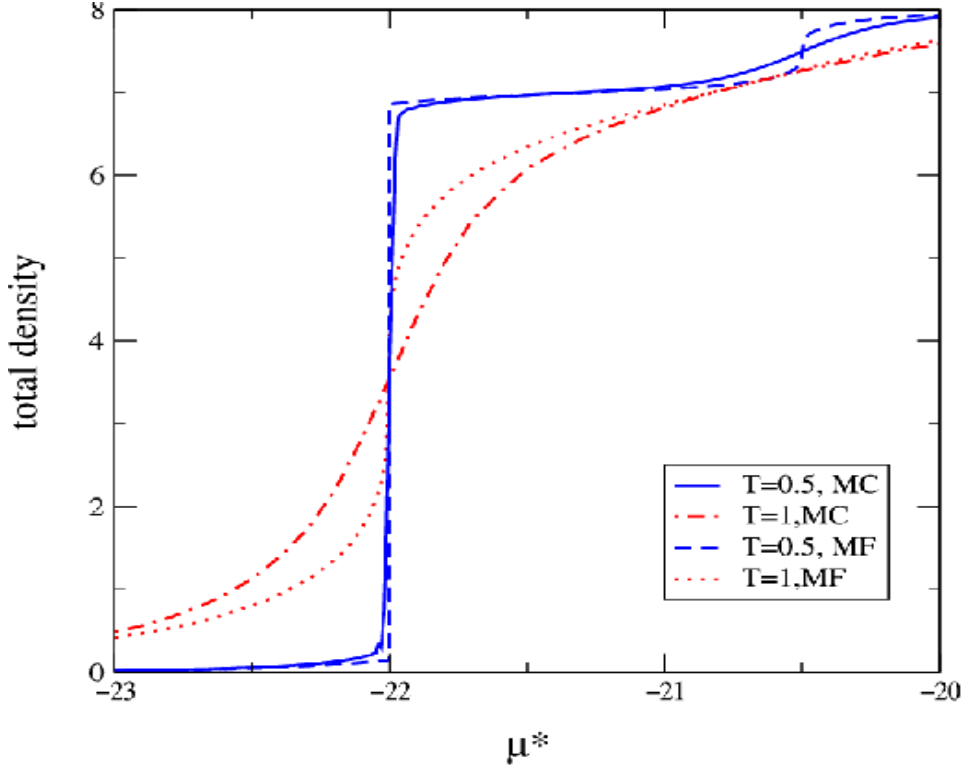}
\vspace*{8pt}
\caption{Mean field isotherms compared with exact Monte Carlo isotherms as a
function of reduced chemical potential and temperature, from
Ref.~\cite{and}.}
\label{fig:isotherms}
\end{figure}

Fig. \ref{fig:isotherms} presents adsorption isotherms for uptake in a
single
cylindrical pore, described by a lattice gas model.\cite{and} In this
model, continuous
space is discretized into adsorption sites, which may be either occupied or
empty in any microstate of the system. In the specific model used here, the
set
of sites (within one transverse section) consists of one axial site and
seven
``cylindrical shell'' sites, corresponding to positions near the inner
boundary
of the nanotube. The ensemble of sites include an infinite sequence of such
layers of eight sites. The energy of the system includes interactions
between
particles occupying these sites, plus interactions between particles and
the
substrate host.

Fig. \ref{fig:isotherms} compares (numerically) exact results with those
obtained from MFT. The results are quite similar, overall, given the highly
expanded scale of reduced chemical potential ($\mu^*$). Note that the
spurious
transition seen in MFT (at reduced temperature $T=0.5, \mu^*=-22$) is not
very
different from the nearly discontinuous isotherm seen in the exact results.

\begin{figure}[bt]
\includegraphics[width=5in,height=3.5in]{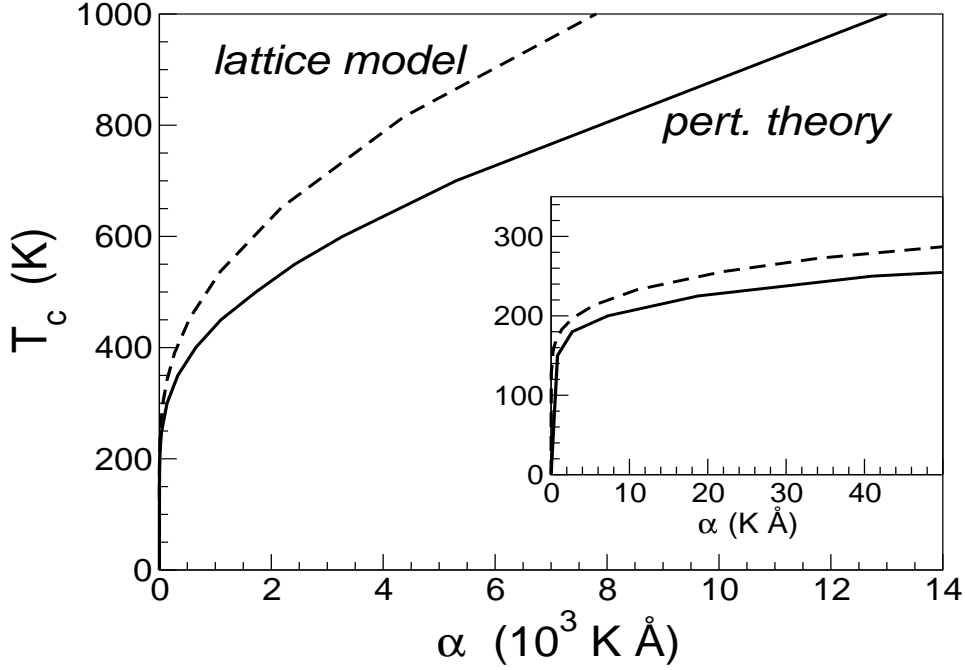}
\vspace*{8pt}
\caption{Critical temperature $T_c$ as a function of the transverse interaction.
Perturbation theory results and predictions from the anisotropic lattice 
gas model are compared. From Ref.~\cite{bucky2}.}
\label{fig:Tc}
\end{figure}

\section{Real transitions of gases within weakly coupled Q1D channels}

Consider a geometry consisting of a set of parallel Q1D fluids, as in
gases 
within or between nanotubes comprising a bundle of such tubes. While no 
transition can occur for an \emph{isolated} Q1D system, once coupling
between such 
systems is present, a finite temperature transition can occur. Fig.
\ref{fig:Tc} presents 
results for such a geometry.\cite{bucky2} The critical temperature $T_c$ is
shown as a function 
of the transverse coupling $\alpha$, for the case of buckyballs confined
within 
parallel nanotubes. As indicated, rather similar results were obtained
from 
exact solutions for a lattice-gas model and from a perturbation theory. In 
the latter case, the  unperturbed equation of state was the exact 1D
result 
for a model with Lennard-Jones interactions between the buckyballs and the 
perturbation was the weak van der Waals interaction between balls in
adjacent tubes.

Particularly striking in this figure is the singular behavior for small
$\alpha$, 
as blown-up in the inset. This behavior is well-known for the lattice-gas,
for 
which the transition temperature satisfies (for weak coupling)\cite{fisher}

\begin{equation}
k_BT_c=\frac{2J_l}{ln(1/c)-ln[ln(1/c)]}
\end{equation}

Here $J_l$ is the longitudinal interaction and $c=J_t/J_l$ is the
anisotropy ratio 
of transverse to longitudinal interaction strengths. The singular behavior 
reflects the divergent susceptibility of the 1D system at low T; as the
correlation 
length diverges, larger regions of fluid in adjacent channels are coupled,
so 
increases rapidly with increasing $J_t$.The other notable result in this
formula 
and the figure is that the characteristic energy and $T_c$ scale are given
by the 
\emph{longitudinal} coupling, which is about 500 K for the peapod case,
whereas naively one might have expected to
be 
proportional to $J_t$.

\section*{Acknowledgments}

We are grateful to DOE and NSF for support of this research and to S. J. Full
and N. Wada for help with figures.

\end{document}